\documentclass[useAMS,usenatbib]{mn2e}
\usepackage{graphicx,amsmath,multirow,amssymb}
\usepackage{subfig}
\usepackage{natbib}
\newcommand{\comment}[1]{}

\def\simgt{\lower.5ex\hbox{$\; \buildrel > \over \F sim \;$}}
\def\simlt{\lower.5ex\hbox{$\; \buildrel < \over \sim \;$}}

\title[Evolved stars in Sextans A]{Evolved stars in the Local Group galaxies - III. 
AGB and RSG stars in Sextans A}
\author[Dell'Agli et al.]{F. Dell'Agli$^{1,2}$, M. Di Criscienzo$^3$, D. A. Garc\'{\i}a--Hern\'andez$^{1,2}$,
P. Ventura$^{3}$, 
\newauthor
M. Limongi$^{3}$, E. Marini$^{3,4}$, O. C. Jones$^{5}$\\
$^{1}$Instituto de Astrof\'{\i}sica de Canarias (IAC), E-38200 La Laguna, Tenerife, Spain \\
$^{2}$Departamento de Astrof\'{\i}sica, Universidad de La Laguna (ULL), E-38206 La Laguna, Tenerife, Spain \\
$^3$INAF -- Osservatorio Astronomico di Roma, Via Frascati 33, 00040, Monte Porzio Catone (RM), Italy \\
$^4$Dipartimento di Matematica e Fisica, Universita degli Studi ``Roma Tre", Via della Vasca Navale 84, I-00146 Roma, Italy\\
$^5$UK Astronomy Technology Centre, Royal Observatory, Blackford Hill, Edinburgh, EH9 3HJ, UK \\
}

\begin{document}

\date{Accepted, Received; in original form }

\pagerange{\pageref{firstpage}--\pageref{lastpage}} \pubyear{2012}

\maketitle

\label{firstpage}

\begin{abstract}
We study the evolved stellar population of the galaxy Sextans A. This galaxy
is one of the lowest metallicity dwarfs in which variable asymptotic giant branch stars
have been detected, suggesting that little metal enrichment took place during the past 
history. The analysis consists in the characterization of a sample of evolved stars, 
based on evolutionary tracks of asymptotic giant branch and red 
super giant stars, which include the description of dust formation in their winds. Use
of mid-infrared and near-infrared data allowed us to identify carbon-rich sources, stars
undergoing hot bottom burning and red super giants. The dust production rate,
estimated as $6\times 10^{-7} M_{\odot}/$yr, is dominated by $\sim 10$ carbon stars, 
with a small contribution of higher mass M-stars, of the order of 
$4\times 10^{-8} M_{\odot}/$yr. The importance of this study to understand how dust production 
works in metal-poor environments is also evaluated.
\end{abstract}

\begin{keywords}
Stars: abundances -- Stars: AGB and post-AGB
\end{keywords}

\section{Introduction}
The research focused on the evolutionary properties of evolved stars have recently
received a renewed interest to inspect their impact on the life cycle of their host galaxy. The DUSTiNGS survey \citep{boyer15a} represents a crucial tool for this scope.
Specifically designed to identify stars evolving through the Asymptotic Giant
Branch (AGB) and massive stars in the core helium-burning phase, this survey provided [3.6] 
and [4.5] magnitudes data collected
with the IRAC camera mounted onboard of the \emph{Spitzer Space Telescope}, for
thousands of stars distributed among 50 dwarf galaxies within 1.5 Mpc.

The analysis of these data will allow to extend to the galaxies in the Local Group (LG)
the studies aimed at the interpretation of IR observations of stellar populations, so far 
limited to the Magellanic Clouds (MC). This will allow testing our 
understanding of the late evolutionary phases to different environments, given the 
diversity in terms of structural properties and star formation histories which 
characterizes LG galaxies.

An important issue regarding the feedback from these stars on the host system is 
the dust formed in the circumstellar envelope and ejected into the interstellar medium.
This is crucial to study the dust evolution in galaxies of the LG 
\citep{dwek98, calura08, schneider14} and for a series of still open points in modern 
Astrophysics, such as the interpretation of the spectral energy distribution (SED) of high-redshift quasars \citep{bertoldi03, wang13} and  the possible explanations of the 
presence of dust at early epochs \citep{valiante11, pipino11}. On this regard, the key 
quantity is the evolution of the dust production rate (DPR) of the stars during 
their life, i.e. the mass of dust produced per unit time in a given evolutionary stage. 

The studies aimed at understanding the dust formation process in the winds of AGB stars
have made significant steps forward in the recent years, when the codes used to model
the AGB evolution were complemented with the equations governing the formation and
growth of dust particles \citep{ventura12a, ventura12b, ventura14, nanni13, nanni14}.

This allowed the study of the evolved stellar populations of the MCs,
characterizing the individual sources observed in terms of mass, chemical composition,
formation epoch and DPR \citep{flavia14, flavia15a, flavia15b, nanni16, nanni18}. To this
purpose, the \emph{Spitzer} mid-IR data were compared with the results from stellar 
evolution + dust formation modelling. The same approach was recently extended
to IC 1613 \citep{flavia16} and IC 10 \citep{flavia18}. The latter studies showed that the 
lack of $[5.8]$, $[8.0]$ and $[24]$ data in the DUSTiNGS survey can be partly compensated
by considering at the same time the $[3.6]$ and $[4.5]$ fluxes with near infrared (NIR) 
observations, providing {\it J-H-K} photometry.

The introduction of this new methodology to investigate the evolutionary properties and
the mechanisms of dust production by stars evolving through the final stages is of
paramount importance, as it represents an alternative and complementary approach to the studies which interpret the IR data with synthetic spectra \citep[see e.g.][]{srinivasan09, srinivasan10, srinivasan11,
boyer11, sargent11, mcdonald12, boyer15c, kraemer17, mcdonald17}. 
The comparison among the results obtained with the two different methods offers a 
valuable opportunity to understand the robustness and the weakness of either approaches, 
and to a more reliable characterization of the stellar populations investigated.

In this paper we study the evolved stellar population of Sextans A, following the same approach
that we adopted for IC 1613 and IC10 by \citet{flavia16, flavia18}. Sextans A is a very peculiar, 
metal-poor dwarf galaxy ($[\rm Fe/H]=-1.4$, Dolphin et al. 2003), with a main stellar component, 
formed between 
1 and 2.5 Gyr ago \citep{weisz14}, and a younger population, formed during the burst in 
the star formation history, which occurred over the past 60 Myr \citep{camacho16}. 
This galaxy is extremely interesting, because it was exposed to little (or no) metal 
enrichment during the last 10~Gyr: it is one of the lowest metallicity objects harbouring
variable AGB stars \citep{boyer15c, mcquinn17}. The study is of paramount importance
to understand how dust production works in metal-poor environments and, more generally,
to clarify whether stars of intermediate mass might play any role in the production of
the large amounts of dust observed at high redshift \citep{gobat18}.

This work is extremely timely, considering that the mid-IR data of Sextans A stars obtained
by DUSTiNGS have been recently completed by near-IR observations by \citet{jones18}, who
present a wide exploration of the stellar population of this galaxy, with the determination
of the individual and global DPR. The study of this galaxy proves extremely important to
test the dust formation theories in metal-poor stars and, on a more general side, to 
assess the reliability of the classification of evolved stars based on a multiwavelength 
approach and the DPR estimates based on SED fitting.

The paper is structured as follows: the sample of Sextans A stars used in the present
work is described in section \ref{sample}; the codes used to model the evolution of AGB 
and RSG stars and the dust formation process are described in section \ref{ATON}; in section 
\ref{models} we discuss the main evolutionary and chemical properties of the evolved
stars of Sextans A; section \ref{distance} is devoted to the characterization of the
sources in the \citet{jones18} sample, whereas in section \ref{dpr} we discuss the DPR of the
individual stars and the global DPR of the galaxy; the effects of a possible metallicity
spread of the evolved stars are discussed in section \ref{zeta}; the conclusions are given
in section \ref{conclusions}.

\section{IR observations of Sextans A}
\label{sample}
To describe the evolved stellar populations of Sextans A we
compare results from stellar evolution computations with
the near IR observations presented recently by  \citet{jones18}.
The  \citet{jones18} sample is made up of 492 stars,
for which JHK photometry was obtained during two observing runs, using the WIYN 
High-resolution Infrared Camera mounted on the $3.5$ m WIYN telescope, at the Kitt Peak
National Observatory. The selected objects are located in
the inner $3'.4 \times 3'.4$ region of Sextans A. This sample is particularly suitable to the main 
goal of the present work, because most  of the sources
with poor signal-to-noise ratio were excluded via a $65 \%$ cut
on the error.  In addition, for a source to be maintained in the
catalogue, it was required that it was detected in two bands (299 objects)
or at two epochs with high confidence.  In general, if a source was detected over multiple 
epochs, the catalogue reports the mean flux for the combined epoch. The completeness of the 
sample is limited by the distance to Sextans A and the weather conditions when the 
observations were taken. In the $K$ band the hystogram of the number of sources becomes 
zero at magnitudes fainter than $19.2$, about half magnitude brigther the tip of RGB. 

In order to fully investigate the nature of the dusty sources, the WHIRC near-IR catalogue  
was matched with the DUSTING catalogue \citep{boyer15a}, adopting a matching distance of 
$3''$; we found 272 objects, 9 out of which were identified as AGB variable stars by 
\citet{boyer15b}.
Stars were classified as O- or C-rich stars according to the colors cut criteria used by 
\citet{cioni06} and \citet{boyer11} in the $K$ versus $J-K$ CMD. They also identified 
dusty AGB stars as those with $J - [8.0] \geq 3.4$ mag or $J - Ks \ge 2.2$ mag 
\citep{blum06,boyer11}. To determine the dust production rate (hereafter DPR), they fit 
the broadband spectral energy distribution with the grid of GRAMS radiative-transfer 
models \citep{sargent11,srinivasan11}. 

\section{AGB and RSG modelling}
\label{ATON}
Considering the low metallicity of stars in Sextans A \citep{dolphin03}, we mostly base 
our analysis on models of stars with mass in the range $1-8~M_{\odot}$ and metallicity 
$Z=10^{-3}$, calculated with the {\sc ATON} code for stellar evolution 
\citep{ventura98}\footnote{$Z=10^{-3}$ corresponds to $[\rm Fe/H]=-1.4$ with the $\alpha-$enhancement
used here, namely $[\alpha/\rm Fe]=+0.4$}.
The evolutionary sequences were started from the pre-main sequence and followed
until the final AGB stages, when the convective envelope was almost entirely consumed.

An exhaustive discussion of the numerical structure of the code {\sc ATON}, with all the 
micro-physics input used, can be found in \citet{ventura98}; the latest updates are given 
in \citet{ventura09}. The choices regarding the macro-physics adopted, particularly for 
what attains the description of the convective instability, are given in \citet{flavia16}.

Regarding the description of mass loss, for oxygen-rich stars we use the
\citet{blocker95} treatment, according to which $\dot M \propto RL^{3.7}/M^{3.1}$; 
the parameter entering the \citet{blocker95}'s recipe is set to $\eta=0.02$, following the 
calibration given in \citet{ventura00}. For carbon stars we used the description of mass 
loss from the Berlin group \citep{wachter02, wachter08}.

To describe dust production for these stars, we apply the description of the wind
of AGB stars proposed by the Heidelberg group \citep{fg06}, which allows to determine
the size of the solid particles formed in the circumstellar envelope, for the various
dust species considered: based on thermal stability arguments, we consider formation of 
silicates and alumina dust in the case of M stars, whereas we assume that solid carbon and 
silicon carbide are the main species produced in the wind of carbon stars. 

\begin{table*}
\caption{The main properties of the stellar models of metallicity $Z=10^{-3}$ and mass 
$0.9~M_{\odot} \leq M \leq 7.5~M_{\odot}$ used in this work. The different columns report
the initial mass of the star (1), the duration of the core H-burning (2) and the AGB 
phases (3), the number of thermal pulses experienced during the AGB evolution (4), 
the maximum luminosity reached during the AGB phase (5), the core mass at the beginning of
the AGB phase (6), the final surface mass fractions of $^{12}$C (7) and $^{16}$O (8),
the final $\log(C-O)+12$ (9, only for carbon stars) and the maximum optical depth at 
$10\mu$m reached (10).
}
\begin{tabular}{c c c c c c c c c c}        
\hline       
$M/M_{\odot}$ & $\tau_H$ (Myr) & $\tau_{\rm AGB}$ (kyr) & NTP & $L_{\rm max} (10^3L_{\odot})$ & 
$M_{\rm core}/M_{\odot}$ & X(C) & X(O) & $\log(C-O)+12$ & $\tau_{10}$ \\
\hline  
 0.9  &  8000  & 1100 &   3   & 3.0  & 0.51 &  7.9e-5  &  5.7e-4 &  -   & 0.0       \\
 1.0  &  5500  & 1750 &   5   & 4.6  & 0.52 &  5.6e-4  &  6.0e-4 & 7.09 & 0.011     \\
 1.1  &  3900  & 1560 &   6   & 5.4  & 0.52 &  3.8e-3  &  8.7e-4 & 8.54 & 0.058     \\
 1.25 &  2500  & 1570 &   8   & 6.0  & 0.53 &  3.7e-3  &  1.0e-3 & 8.51 & 0.065     \\
 1.3  &  2200  & 1590 &   8   & 6.5  & 0.53 &  6.6e-3  &  1.3e-3 & 8.79 & 0.096     \\
 1.5  &  1450  & 1600 &   9   & 8.0  & 0.54 &  1.5e-2  &  2.2e-3 & 9.17 & 0.15      \\
 1.75 &  980   & 1350 &   11  & 10   & 0.56 &  1.6e-2  &  4.2e-3 & 9.15 & 0.26      \\
 2.0  &  700   & 2000 &   17  & 11   & 0.56 &  2.5e-2  &  2.5e-3 & 9.41 & 0.54      \\
 2.25 &  530   & 1380 &   18  & 13   & 0.60 &  2.1e-2  &  1.0e-2 & 9.18 & 0.55      \\
 2.5  &  410   & 860  &   18  & 15   & 0.66 &  1.6e-2  &  4.2e-3 & 9.15 & 0.56     \\
 3.0  &  270   & 330  &   25  & 25   & 0.79 &  2.6e-3  &  1.6e-3 & 8.19 & 0.43      \\ 
 3.5  &  197   & 300  &   29  & 31   & 0.81 &  3.8e-4  &  9.2e-4 &  -   & 0.37      \\
 4.0  &  142   & 230  &   36  & 40   & 0.85 &  2.1e-4  &  3.9e-4 &  -   & 0.56      \\
 4.5  &  112   & 170  &   38  & 48   & 0.88 &  1.0e-4  &  2.2e-4 &  -   & 0.59      \\
 5.0  &  90    & 126  &   42  & 57   & 0.92 &  1.0e-4  &  3.7e-4 &  -   & 0.64      \\
 5.5  &  74    & 59   &   31  & 70   & 0.97 &  2.0e-5  &  1.5e-5 &  -   & 1.3       \\
 6.0  &  62    & 45   &   38  & 89   & 1.02 &  1.6e-5  &  1.1e-5 &  -   & 1.68      \\
 6.5  &  53    & 40   &   31  & 107  & 1.12 &  1.9e-5  &  2.7e-5 &  -   & 1.69      \\
 7.0  &  47    & 17   &   28  & 121  & 1.19 &  2.0e-5  &  4.1e-5 &  -   & 1.695     \\
 7.5  &  41    & 11   &   21  & 150  & 1.28 &  2.4e-5  &  1.6e-4 &  -   & 1.71      \\ 
\hline       
\label{tabmod}
\end{tabular}
\end{table*}

To calculate the dust produced by the stars during their life, we consider several points
selected during the AGB phase and we model dust formation on the basis of the
corresponding values of mass, mass loss rate, luminosity, effective temperature and
the surface chemical composition. This is the method which we used in a series of papers
\citep{ventura12a, ventura12b, ventura14}, to calculate the dust expected from AGB stars.

Finally, the knowledge of the aforementioned physical quantities and of the kind and
size of the dust particles formed allow the calculation of the synthetic spectra and 
the determination of the IR fluxes; for the latter step we use 
the code {\sc DUSTY} \citep{dusty}.

Consistently with the previous works on this argument, we will quantify the degree
of obscuration of the stars by the optical depth at $10 \mu$m, $\tau_{10}$, which is
determined by the radial distribution of the dust particles in the surroundings of the star
(see Section 2.3 in Dell'Agli et al. 2015a).

The interpretation of the observations in the colour-magnitude diagrams must consider the
presence of RSG stars, evolving through the core helium-burning phase. For the stars
with mass $M < 8~M_{\odot}$ we considered the models discussed above, calculated with the
{\sc ATON} code.
For higher mass stars we considered stellar models of metallicity $[\rm Fe/H]=-2$,
calculated by means of the rotating version of the FRANEC code\footnote{Note that the
FRANEC models are available for $[\rm Fe/H]=0$, $-1$, $-2$, $-3$} (the interested reader is 
referred to \citet{chieffi13} for more details).

\section{Stars evolving in Sextans A: physical and chemical properties}
\label{models}
The vast majority of the evolved stellar population of Sextans A is composed by 
stars evolving through the AGB phase, the progeny of stars with mass in the range 
$0.9~M_{\odot} \leq M \leq 7.5~M_{\odot}$. Stars with mass above $7.5~M_{\odot}$ undergo 
core collapse, thus they do not experience the AGB phase. We will not consider stars of 
mass below $0.9~M_{\odot}$, because, independently of the description of mass loss during 
the red giant branch (RGB) phase, these stars will never evolve at luminosities above the 
tip of the RGB and will not produce any kind of dust. For the reasons discussed in the 
previous sections, we focus on a metal-poor chemical composition, with metallicity 
$Z=10^{-3}$, though a little spread in $Z$ cannot be excluded. We will explore this 
possibility in section \ref{zeta}.

The properties of AGB stars are described in excellent reviews on the
argument. The interested reader is addressed to the works by \citet{busso99}, \citet{herwig05},
\citet{karakas14} for a thorough discussion of the main physical mechanisms affecting the
AGB evolution, the processes able to alter the internal and the surface chemical composition
and the most relevant uncertainties in the micro- and macro-physics input adopted.

Tab.~\ref{tabmod} reports the main properties of the models used here. The luminosity
experienced during the AGB phase, reported in col. 5, increases with the initial mass
of the star, because the higher the mass the higher the core mass at the beginning
of the AGB phase (see col. 6). This is the reason why the overall duration of the AGB
phase, reported in col. 3, gets shorter for larger values of the initial mass.
The number of thermal pulses (TP) experienced first increases with the initial mass, 
because higher mass
stars loose the envelope after a higher number of TP; this trend is reversed for stars of
mass above $\sim 5~M_{\odot}$, because in this mass domain the rate of mass loss becomes
so large to favour a fast loss of the envelope, with a smaller number of TP. 

\subsection{The surface chemical composition}
\begin{figure*}
\begin{minipage}{0.48\textwidth}
\resizebox{1.\hsize}{!}{\includegraphics{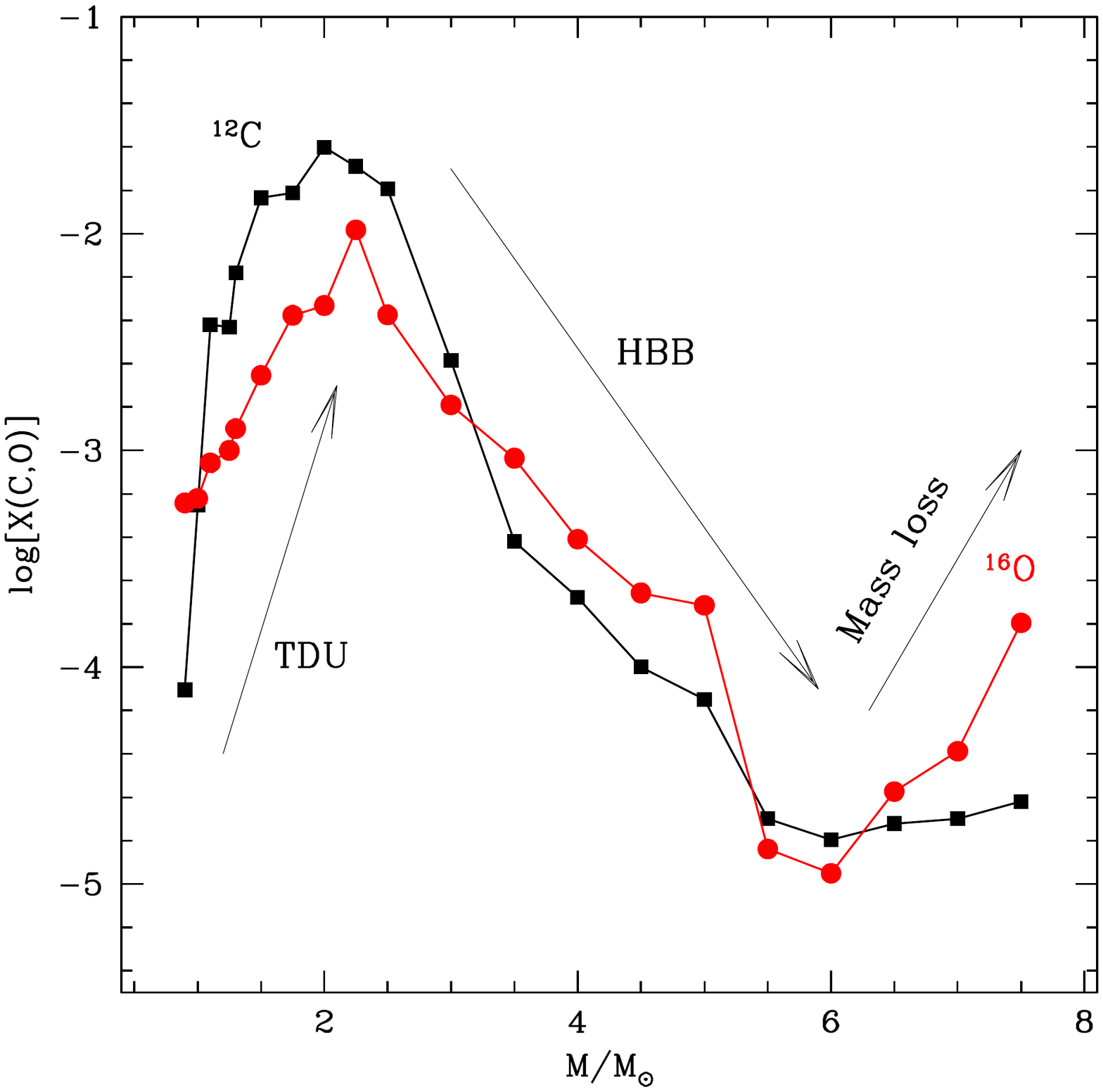}}
\end{minipage}
\begin{minipage}{0.48\textwidth}
\resizebox{1.\hsize}{!}{\includegraphics{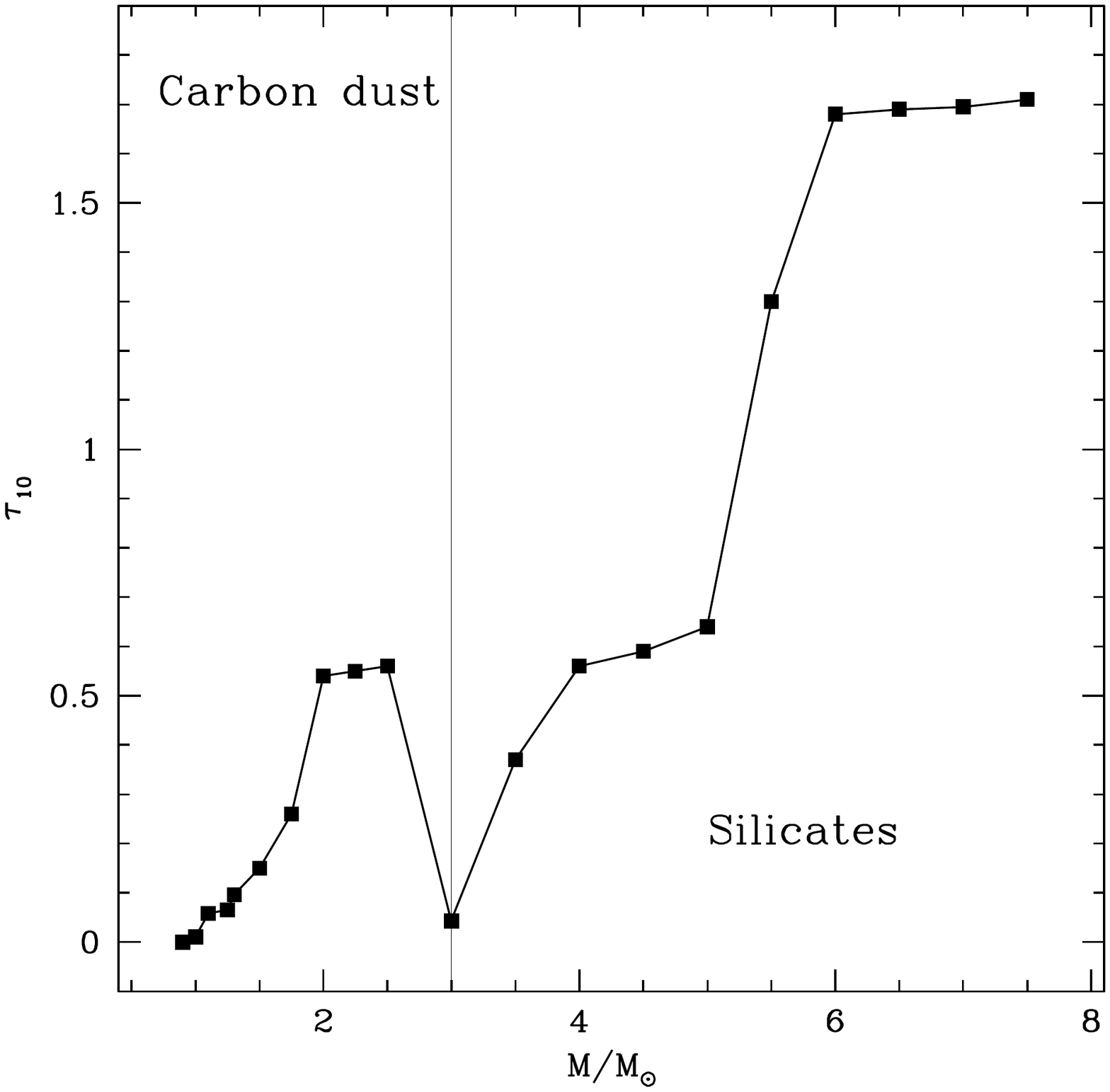}}
\end{minipage}
\vskip-50pt
\caption{Left: The final surface mass fractions of carbon (black squares) and
oxygen (red points) for the $Z=10^{-3}$ models with mass $0.9 \leq M/M_{\odot} \leq 7.5$
discussed in the text. The effects of TDU, HBB and the enhanced mass loss in massive
AGB models are indicated with arrows (see text for details). Right: The maximum optical 
depth reached during the AGB phase. The line separates low mass stars which form mainly carbon dust and stars where silicates are formed predominantly.}
\label{fmodels}
\end{figure*}

An important point for the present work is the modification of the surface chemical
composition as the stars evolve through the AGB. The left panel of 
Fig.~\ref{fmodels} shows the surface abundances of carbon and oxygen at the
end of the AGB phase, as a function of the initial mass. It is clear in the
figure the different behaviour followed by stars of different mass:

\begin{enumerate}

\item{For stars of mass below $3~M_{\odot}$ the only mechanism active in changing
the surface chemical composition is third dregde-up (hereafter TDU), which
favours a significant enrichment in the surface carbon and a small increase in the
oxygen content; eventually carbon replaces oxygen as the dominant species in the envelope,
making the star to become a carbon star. Fig.~\ref{fmodels} shows that the carbon enhancement
increases with the stellar mass in this mass domain, because the higher the initial mass the
higher the number of TDU episodes experienced
(see col.~4 in Tab.~\ref{tabmod}). This trend changes when the mass approaches the 
upper limit of the stars which attain the C-star stage.}

\item{The stars of mass $M > 3~M_{\odot}$ experience hot bottom burning (HBB) during the
AGB phase \citep{renzini81, blocker91}. The base of the convective envelope is exposed
to severe p-capture nucleosynthesis, which destroys the surface carbon and, in metal
poor environments, also
the oxygen nuclei. The final carbon and oxygen mass fractions shown in Fig.~\ref{fmodels}
decrease with the initial mass of the star, because models of higher mass evolve on
more massive cores and suffer a stronger HBB nucleosynthesis \citep{ventura09}. The 
final oxygen in the models of mass above $6~M_{\odot}$ is higher than in their lower 
mass counterparts, because mass loss is so fast that the envelope is lost before a very 
advanced degree of nucleosynthesis is reached \citep{ventura11}.}

\end{enumerate}

\subsection{Dust production}
\label{dustmod}
Understanding the surface chemistry variation of AGB stars proves important for 
the dust formed in their circumstellar envelope and ejected into their
surroundings. The results shown in the left panel of Fig.~\ref{fmodels} allow us to
deduce the main dust species formed for stars of different mass. The main argument
here is that the CO molecule is extremely stable: in C-rich environments no dust species
which include oxygen particles can be formed, whereas in oxygen-rich envelopes there are
no C-bearing molecules to form carbonaceous grains \citep{fg06}. Therefore, we find that
low-mass stars produces mainly carbon dust, whereas the stars experiencing HBB form
only silicates and alumina dust. The behaviour of the dust formed with the initial mass of
the stars is thoroughly discussed in \citet{ventura14}.

The right panel of Fig.~\ref{fmodels} shows the quantità of dust produced
by the AGB models used in the present investigation; the quantity reported on the y-axis
is the maximum optical depth (at $\lambda = 10~\mu$m) reached during the AGB phase, which
is sensitive to the radial distribution and the size of the dust grains present in the
circumstellar envelope.

In the low-mass domain ($M < 3~M_{\odot}$) $\tau_{10}$ increases with the initial mass of
the star. This is not surprising, based on the behaviour of the 
surface carbon, shown in the left panel of Fig.~\ref{fmodels}: models of higher mass 
accumulate a larger amount of carbon in the surface regions, which favours a faster growth 
of the newly formed solid carbon particles, hence higher values of $\tau_{10}$.  
The largest $\tau_{10}$ are reached during the very final AGB phases, 
when the surface carbon is largest \citep{flavia15a,ventura16b}. 
The contribution of SiC to the overall dust
produced can be neglected: for this metallicity the fraction of silicon in the envelope 
is so small that the amount of SiC is a factor of 10 or more smaller than solid carbon
\citep{ventura14}.

For the stars experiencing HBB ($M > 3~M_{\odot}$), the dust formed is under the form of
silicates and alumina dust (Al$_2$O$_3$). The latter species is very stable and forms in 
a more internal region of the circumstellar envelope ($2-3$ stellar radii form the 
surface) compared to silicates ($5-10$ stellar radii). On the other hand, the amount of 
aluminium in the star is much smaller than silicon and Al$_2$O$_3$ is extremely 
transparent to the electromagnetic radiation: therefore, we may safely assume that the 
reprocessing of the radiation is exclusively due to silicate particles \citep{flavia14}.

These stars attain extremely large luminosities and experience high rates of mass loss: 
these conditions prove extremely favourable to the formation and growth of dust particles, 
because large mass loss rates favour the formation of extremely dense winds, with a larger 
availability of gas molecules to form dust. Fig.~\ref{fmodels} shows that $\tau_{10}$ 
increases with the mass of the star, because stars of higher mass evolve at larger 
luminosities (see Tab.~\ref{tabmod}) and suffer very strong mass loss, with rates above 
$10^{-4} M_{\odot}/$yr. 

Before discussing how dust production affects the colours of evolved M-star
in the observational planes, we believe important to stress here that the above results,
concerning the formation of silicates, 
are based on the simplified model for the growth of dust particles, described in section 
\ref{ATON}. This model assumes that the wind is stationary and, more important, the
rate of mass loss is assumed a priori, thus independent of dust formation. This is the
only possible approach within the current modelling of dust from AGB stars, if coupling
with an evolutionary description is required.
Detailed hydrodynamical simulations showed that silicate particles formed at the distances
from the surface of the star found in the present work can hardly drive a wind
\citep{hofner08, bladh12}. These results refer to solar mass stars though, thus the
extension to the case of the massive AGB stars discussed here, evolving at mach higher
luminosities, is not straightforward. In the present analysis we find that the 
radiative push exceeds the gravitational pull and is sufficient to accelerate the wind,
up to velocities in the range $10-20$ Km/s. However, the incoming hydrodynamical models, 
with mass in the range of interest here, will definitely
clarify whether the conditions to radiatively drive a wind are shared by the present
models.

\subsection{Evolutionary tracks in the observational planes}
\label{sectracks}
\begin{figure*}
\begin{minipage}{0.48\textwidth}
\resizebox{1.\hsize}{!}{\includegraphics{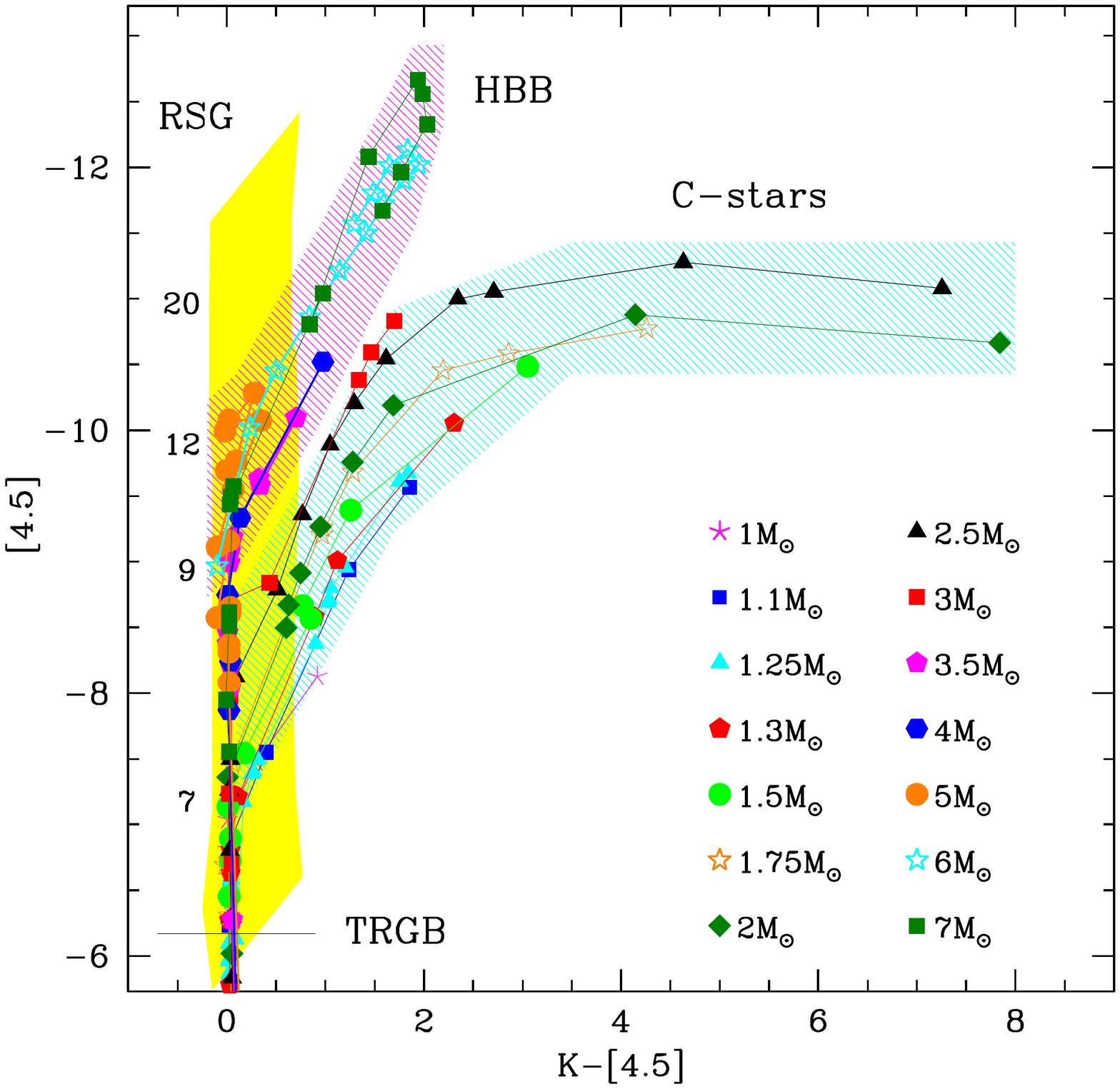}}
\end{minipage}
\begin{minipage}{0.48\textwidth}
\resizebox{1.\hsize}{!}{\includegraphics{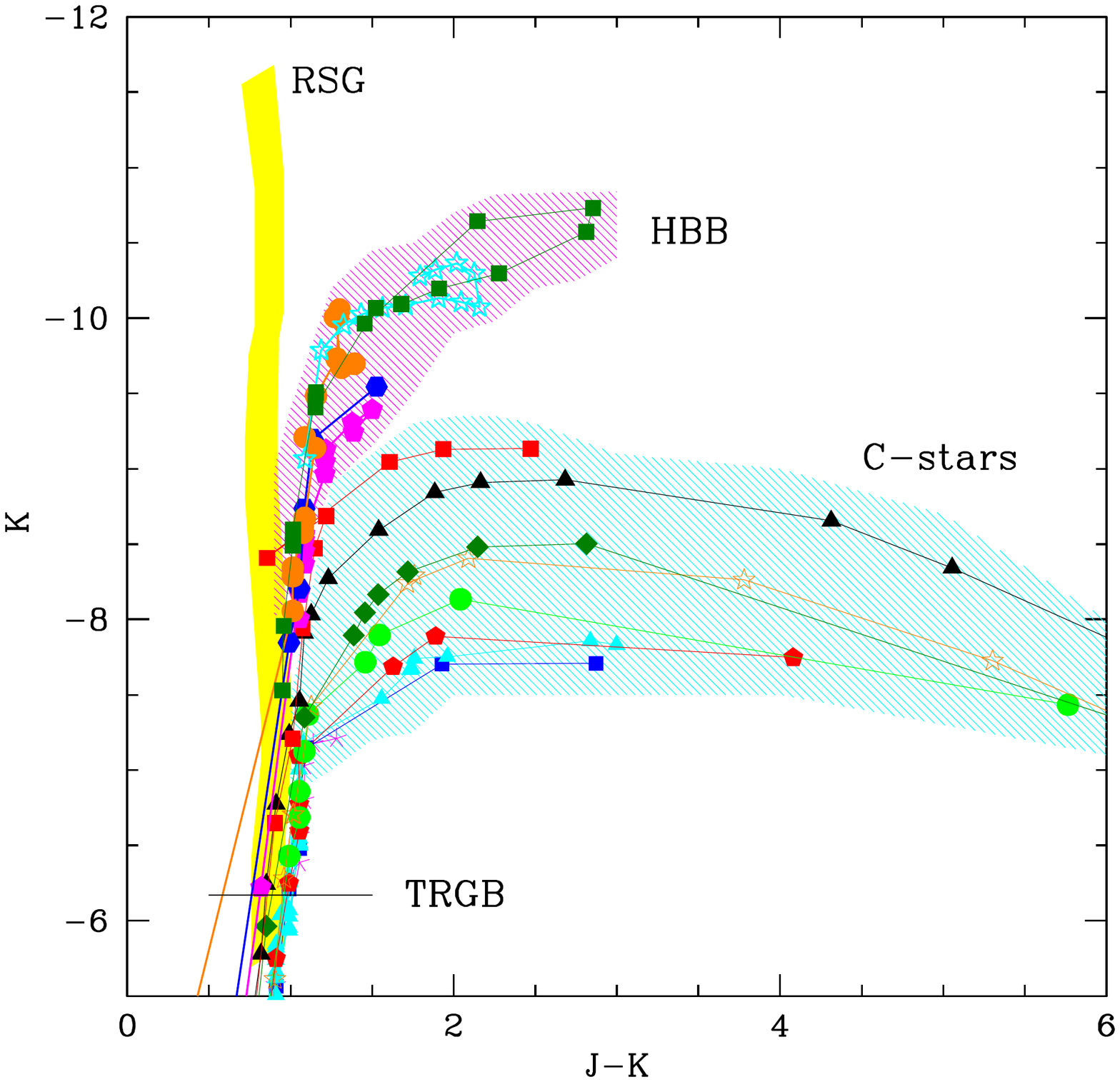}}
\end{minipage}
\vskip-50pt
\caption{The evolutionary tracks of $Z=10^{-3}$ stars of different mass in the 
colour-magnitude $(K-[4.5], [4.5])$ (left) and $(J-K, K)$ (right) planes. The regions
highlighted with cyan and magenta shading indicate, respectively, the zones
where carbon stars and intermediate mass stars, experiencing HBB, evolve. The yellow
region indicate the expected position of core helium-burning stars; for the latter
an artificial spread in the optical depth of $\delta \tau_{10} = 0.05$ was assumed
(see text for details).}
\label{ftracks}
\end{figure*}

Fig.~\ref{ftracks} shows the evolutionary tracks of the AGB models used in the present
analysis, in the colour-magnitude $(K-[4.5],[4.5])$ and $(J-K, K)$ planes
(hereafter CMD1 and CMD2, respectively). 

Low mass stars ($M \leq 3~M_{\odot}$) increase the degree of obscuration as 
they evolve through the AGB, owing to the gradual accumulation of 
carbon in the surface regions. In the CMD1 the evolutionary tracks first move to the red,
following a diagonal path, with higher and higher $[4.5]$ fluxes. This behaviour holds 
as far as $K-[4.5] < 2$. For redder colours the tracks flatten, because the peak in the SED
gets close and eventually overcomes the wavelengths covered by the $[4.5]$ 
filter \citep[see e.g. Fig.~6 in][]{flavia15a}, thus no further increase in the $[4.5]$ 
flux occurs.

Carbon stars populate the region in the CMD1, highlighted with a cyan 
shading in the left panel of Fig.~\ref{ftracks}, which extends to $K-[4.5] \sim 8$. Only 
stars of initial mass $M > 1.5-2~M_{\odot}$ are expected to reach colours $K-[4.5] > 3$, 
in agreement with the earlier discussion and the results shown in Fig.~\ref{fmodels}. 
Regarding the region $K-[4.5] < 3$, we expect for a given ($K-[4.5]$) a spread in $[4.5]$
of approximately 1 mag, with the stars of higher mass being brighter.
 
Stars of initial mass $M > 3~M_{\odot}$ reach larger luminosities during the AGB phase
(see Tab.~\ref{tabmod}), thus the evolutionary tracks lay above those corresponding to
their smaller mass counterparts. The obscuration sequence of these stars, indicated with 
a magenta box in Fig.~\ref{ftracks}, extends to $K-[4.5] \sim 2$, and is separated
by the C-star zone. Only massive AGB stars of initial mass above $6~M_{\odot}$ are
expected to evolve to the red side of the CMD1, because for lower masses, considering that
we are discussing metal-poor chemistries, little dust production 
takes place in the envelope; this is consistent with the drop in the values of $\tau_{10}$
below $6~M_{\odot}$, shown in the right panel of Fig.~\ref{fmodels}. While in C-stars the 
main factor affecting dust formation 
(and the consequent reddening of the evolutionary track) is the increase in the surface 
carbon \citep{ventura16b}, in this case significant amounts of dust are formed under 
strong HBB conditions, which favour large rates of mass loss \citep{ventura15}.

The evolutionary paths in the CMD2, shown in the right panel of Fig.~\ref{ftracks}, 
are qualitatively similar to the tracks in CMD1.
The obscuration sequence of carbon stars is more extended in this case, because the
$J$ flux gets extremely faint in the most advanced evolutionary phases, when the radiation
is reprocessed by solid carbon particles in the circumstellar envelope. The higher
sensitivity of $J-K$ to the optical depth might suggest that the CMD2 plane is suitable
to study the obscured stars; however, the almost null values of the $J$ flux makes extremely
hard to predict $J-K$ for these objects and it is common that the flux in the 
$J$ band is not available for the majority of them. This is the reason
why CMD1 proves more useful to study the stars with the largest degree of obscuration.

On the other hand the analysis of the position of the stars on CMD2 is extremely useful
to study the phases following the achievement of the C-star stage, because the
$J-K$ colour is more sensitive to the optical depth when the degree of obscuration is
small, compared to $K-[4.5]$. The same holds for the study of the stars experiencing HBB.

Regarding the stars experiencing HBB, their position on CMD2 proves particular useful
in this case, because, the optical depths reached are generally small, owing to the low
metallicity, and the $J-K$ colour is more sensitive to $\tau_{10}$ for low degree of
obscurations. Indeed we see in Fig.~\ref{ftracks} that all the tracks of $M>3~M_{\odot}$
stars attain $J-K$ colours at least $\sim 0.5$ mag redder than the unobscured stars;
conversely, on the CMD1 only the very massive AGB stars reach an appreciable degree of
obscuration, during the phases with the strongest HBB.

In the CMD1 and CMD2 planes shown in Fig.~\ref{ftracks} we highlight with a yellow
shading the RSG regions, i.e. the zones where stars of mass in the range $5-30~M_{\odot}$ 
evolve during the core He-burning phase. In the left panel we also indicate the position 
of the stars of a given mass on the CMD1. Unfortunately no solid theory regarding dust 
formation in the winds of these stars is currently available. 
Following the same approach used by \citet{flavia18}, we arbitrarily assumed a spread in 
$\tau_{10}$ for RSG stars, with $0 < \tau_{10} < 0.05$. The width of the yellow regions 
in Fig.~\ref{ftracks} correspond to such a spread in $\tau_{10}$\footnote{We use a narrower 
range of $\tau_{10}$ with respect to \citet{flavia18}, because we considered that the 
stars in Sextans A are on the average more metal poor than in IC10.}.

\section{The evolved stellar population of Sextans A}
\label{distance}
To study the evolved stellar population of Sextans A we analyse the position in the 
observational colour-magnitude planes of the stars belonging to the sample by 
\citet{jones18}, based on the evolutionary tracks presented in the previous section. 

We first determine a bona fidae distance of the galaxy, by means of 
the same method applied by \citet{flavia18} for IC10. We shift the evolutionary tracks 
vertically, until achieving overlapping between tracks and observations. 
In particular, we focus on the diagonal sequence extending to the red side of CMD1 and 
CMD2, indicated with a cyan shading in Fig.~\ref{ftracks}, populated by carbon stars.
A reasonable match is obtained for a distance of $d=1.4$ Mpc, in agreement with 
\citet{dolphin03} and \citet{mccon12}.

We base the characterization of the individual stars on the position on the CMD1
and CMD2 planes. For the stars with no $J$ ($[4.5]$) flux available, we rely
on the CMD1 (CMD2) plane only. The other sources, located inside the cyan (magenta) 
regions in both planes, were classified as carbon (HBB) stars. 

\begin{figure*}
\begin{minipage}{0.48\textwidth}
\resizebox{1.\hsize}{!}{\includegraphics{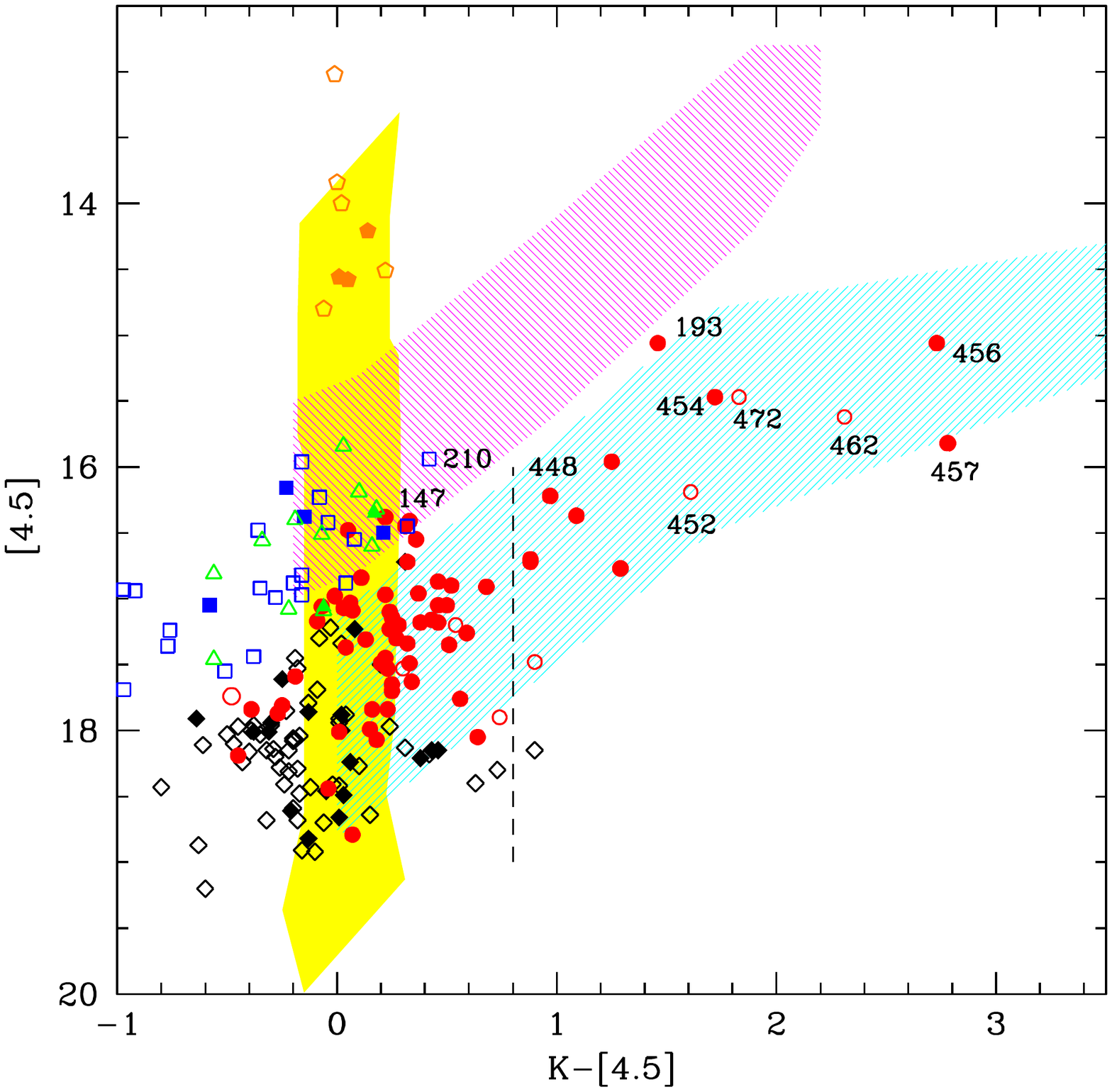}}
\end{minipage}
\begin{minipage}{0.48\textwidth}
\resizebox{1.\hsize}{!}{\includegraphics{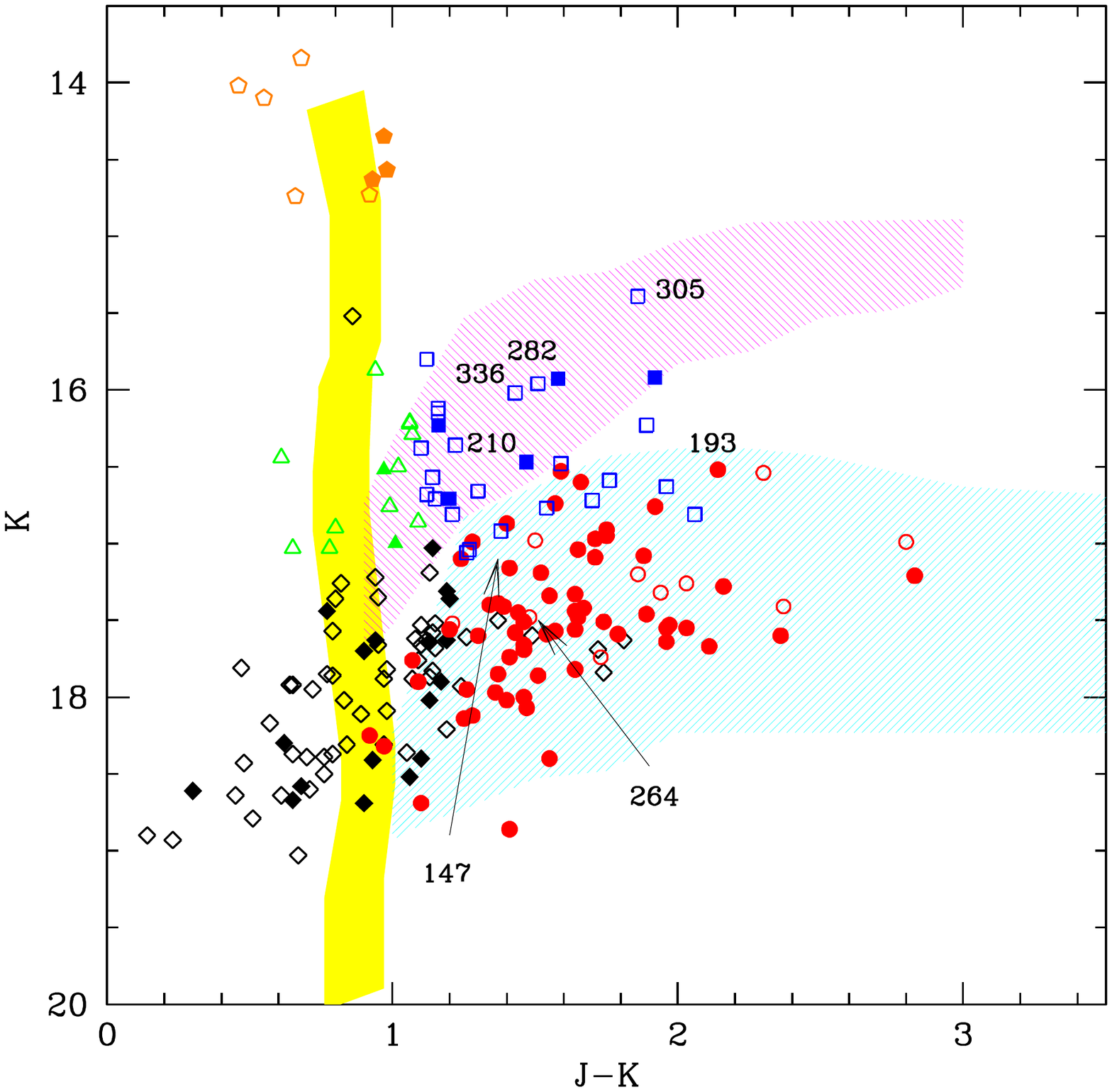}}
\end{minipage}
\vskip-50pt
\caption{The position of evolved stars belonging to Sextans A in the colour
magnitude $(K-[4.5], [4.5])$ (left) and $(J-K, K)$ (right) planes. The meaning of the
various symbols indicate our classification, according to the following coding:
red circles: carbon stars; blue squares: intermediate mass stars, undergoing HBB;
green triangles: HBB stars or RSG stars; orange pentagons: RSG stars; black diamonds:
low-intermediate mass, scarcely obscured stars or core helium burning stars. 
Full (open) points indicate carbon (oxygen-rich) 
stars, according to the classification by \citet{jones18}. The cyan, magenta and
yellow regions indicate the zones in the observational planes where carbon stars,
intermediate mass stars and RSG stars evolve. The numbers indicate the ID classification
in the \citet{jones18} catalogue of some sources of particular interest. The sources 
in the left panel on the right of the vertical, dashed line are C-stars with a very large
degree of obscuration, with $\tau_{10} \geq 0.1$}
\label{foss}
\end{figure*}

For the stars exhibiting a small degree of obscuration, with $J-K$ ($K-[4.5]$) 
only slightly above $\sim 1$ ($\sim 0$), we mainly relied on their position on the CMD2
plane. This choice is motivated by the arguments discussed in section \ref{sectracks}:
the $J-K$ colour, compared to $K-[4.5]$, is more sensitive to the optical depth in 
low $\tau_{10}$ stars. This criterion was mainly used to characterize HBB stars, 
because only a small degree of obscuration of these sources is expected in this galaxy, given the 
small metallicities. In particular, the stars whose position on the CMD2 overlaps with 
the evolutionary tracks of stars of mass above $3~M_{\odot}$ and/or with the region
populated by RSG stars, were classified as HBB or RSG stars, even in the cases, clearly
visible in the left panel of Fig.~\ref{foss}, where their $K-[4.5]$ colours on the CMD1 
are significantly bluer ($K-[4.5]<0$) than the expectations.

We did not classify stars with $J-K < 1.3$ and $K-[4.5] < 0.2$, leaving all
the possibilities open; the faintest among these objects might descend from low-mass
progenitors or from more massive stars currently experiencing a scarce obscuration phase,
whereas the brighter objects might be (weak) HBB or RSG stars.  

The classification proposed here, based on the above arguments, has some degree of 
arbitrariness; we cannot rule out that some of the sources which according to our understanding 
are C-stars or HBB stars were misclassified. However, this uncertainty concerns only
objects with a small degree of obscuration, which provide a negligible feedback on the
host system, and whose contribution to the overall dust production rate of the galaxy
is practically null. This is not going to affect the main conclusions of the present 
investigation.

Before discussing the results obtained we believe important to stress here that testing 
our interpretation will require complete photometry at near-IR and mid-IR wavelengths
with high spatial resolution; this task will be possible, considering the distance of
Sextans A, when the JWST will be finally operative.

We find that $\sim 90$ sources in the \citet{jones18} sample are carbon stars. These 
objects, which fall within the cyan region in Fig.~\ref{foss}, are indicated with red points 
and circles. Following the discussion in section \ref{models}, we conclude that the 
progenitors of these sources were stars of mass in the range $\sim 1-3~M_{\odot}$, formed 
between $500$ Myr and $\sim 6$ Gyr ago (see Tab.\ref{tabmod}). These stars are currently 
evolving at luminosities $5000~L_{\odot} < L < 15000~L_{\odot}$. The repeated TDU
episodes experienced made the surface layers to be enriched in carbon, with mass
fractions in the range $5\times 10^{-4} < X(C) < 0.015$ and carbon excess, defined
according to \citet{lars10}, $\log(C-O)+12 \sim 9.4$.

C-stars trace an obscuration sequence in CMD1 and CMD2. The larger the surface mass 
fraction of carbon, the higher the size of the dust grains formed, the redder the colour.
The circumstellar envelope of these stars hosts solid carbon grains, with size 
$0.03 ~ \mu$m $< a_C < 0.2 ~ \mu$m, and SiC particles, $\sim 0.05 ~ \mu$m sized. 
As discussed in section \ref{dustmod}, the latter kind of dust has negligible effects on the
IR properties of metal-poor stars. The size of the carbon grains formed partly depends 
on the choice of the number density ($n_d$) of seeds particles available. In the present work,
consistently with previous investigations published on this argument, we assumed
$n_d=10^{-13}n_H$. A higher(lower) $n_d$ would favour the formation of smaller (bigger)
grains. A thorough discussion on how these choices reflect into the dimension of the
carbon particles formed is found in \citet{nanni16}.

Among the C-stars group, the most interesting sources are the 14 objects with the largest 
IR excess, with $K-[4.5] > 0.8$. These objects populate the right side of CMD1, the region 
delimited by the dashed, vertical line in the left panel of Fig.~\ref{foss}. Note that for 
only 4 of these stars the $J$ flux is available, which confirms the usefulness of the CMD1 plane 
to study the most obscured C-stars. According to our understanding, in these sources the surface
carbon mass fraction is above $\sim 1\%$, with a carbon to oxygen ratio
$C/O = 5-10$ and a carbon excess $\log(C-O)+12 > 8$. They are surrounded by significant 
quantities of carbon dust, with solid 
carbon grains of dimensions $0.1 ~ \mu$m $< a_C < 0.2 ~ \mu$m. The fractions of gaseous carbon
condensed into dust grains ranges from $5\%$ to $30\%$.

Moving away from the C-star sequence, we identified $\sim 30$ objects, indicated with blue 
squares in Fig.~\ref{foss}, which we interpret as stars currently undergoing 
HBB; the latter mechanism, as discussed in section \ref{dustmod}, favours efficient formation 
of dust in the wind. Most of these sources fall within the magenta box in Fig.~\ref{foss}. 
Six of these objects lay at the edge between the magenta and the cyan boxes in the
CMD2, which makes their classification tricky. We chose to classify these stars as HBB stars,
because their $[4.5]$ fluxes are too large to be C-stars; this classification, which is in
agreement with the conclusions by \citet{jones18}, is however uncertain.

Stars in this group are the progeny of intermediate mass stars, with initial masses
above $\sim 4~M_{\odot}$ and ages younger than $200$ Myr. 
The surface chemical composition of these objects reflects the effects of HBB at the base
of the envelope, with the destruction of carbon by proton capture reactions; 
given the small metallicities, we also expect some oxygen depletion in the external 
regions (Ventura et al. 2013, see also Fig.~\ref{fmodels}). 
The surface carbon mass fraction is $X(C) \sim 2\times 10^{-5}$ 
(note that initially it was $4-5$ times higher), whereas the oxygen abundance is 
$X(O) \sim 5\times 10^{-5}$, a factor of $\sim 10$ smaller than in the gas from which
the stars formed. These stars are extremely bright, with luminosities of the order of
$4-10 \times 10^4~L_{\odot}$ (see tab.~\ref{tabmod}), triggered by the ignition of 
HBB \citep{blocker91}.

As discussed in section \ref{models}, the dust formed in the wind is mainly composed of
silicates, which reach sizes of the order of $0.05-0.07 \mu$m. The fraction of silicon
condensed into dust spans the range $10-50~\%$ \citep{ventura12a}. The formation of alumina 
dust particles, with size $0.03-0.05 \mu$m, is also expected in the circumstellar 
envelope \citep{flavia14, jones14}.

The existence of these stars, descending from progenitor of mass in the range
$4-7.5~M_{\odot}$, is a clue that significant star formation took place in
Sextans A in the last 200 Myr, in agreement with  \citet{dolphin03}.

The \citet{jones18} sample includes a sub-sample of $\sim 20$ stars, with a small degree of
obscuration, whose $K$ and $[4.5]$ fluxes are sufficiently bright to conclude that they 
descend from $M > 3~M_{\odot}$ progenitors. Unfortunately in these regions of the CMD1 and
CMD2 planes the tracks of no-dusty AGB stars and RSG stars are overlapped; therefore,
we leave open the possibilities that these sources, indicated 
with green triangles in Fig.~\ref{foss}, either descend from
$4-8~M_{\odot}$ progenitors, currently experiencing the initial AGB phases, 
with little dust in their envelope, or that they are RSG stars, of mass 
in the range $8-12~M_{\odot}$, currently evolving through the core He-burning phase. 
The analysis of photometric variability or the determination of the chemical composition 
of these objects from high-resolution spectroscopy would allow a straightforward
identification of their nature \citep{garcia17}.

The sample by \citet{jones18} is completed by $\sim 350$ faint objects ($[4.5]>17$), which are 
not expected to provide a significant contribution to the stellar feedback to the galaxy,
particularly for what concerns dust production. The characterization of these stars
is not trivial, considering that only 88 (84) of them are present in the CMD1 (CMD2) plane.
Furthermore, the zones of the observational planes populated by these stars are crossed
by the evolutionary tracks of low-mass ($M \leq 3~M_{\odot}$) stars evolving through AGB, 
intermediate mass stars during the early AGB phases, and $5-8~M_{\odot}$ stars in the core 
helium burning phase. Based on the relative duration of these phases we believe that the
majority of these objects descend from low-mass progenitors 
($0.9~M_{\odot} < M < 3~M_{\odot}$), and are currently experiencing the TP
phase.

The data available do not allow us to rule out that a fraction of these objects has
reached the C-star stage, with a small degree of obscuration. According to \citet{jones18}
$15\%$ of them are carbon stars.

\section{The dust production rate}
\label{dpr}
The characterization of the individual sources belonging to Sextans A, according to the
criteria described in the previous section, allows us to provide an estimate of
the DPR from each star.

\begin{figure*}
\begin{minipage}{0.48\textwidth}
\resizebox{1.\hsize}{!}{\includegraphics{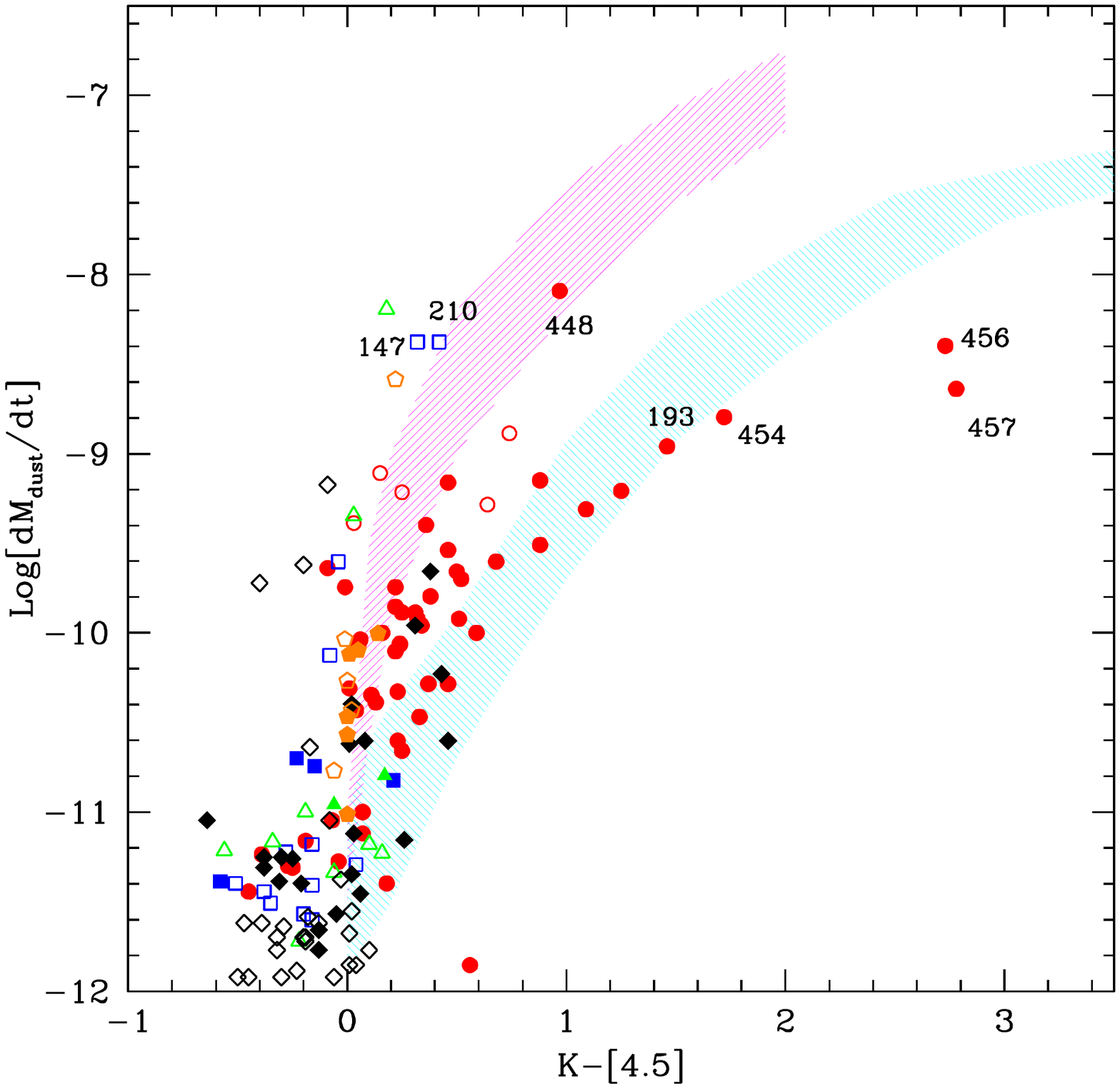}}
\end{minipage}
\begin{minipage}{0.48\textwidth}
\resizebox{1.\hsize}{!}{\includegraphics{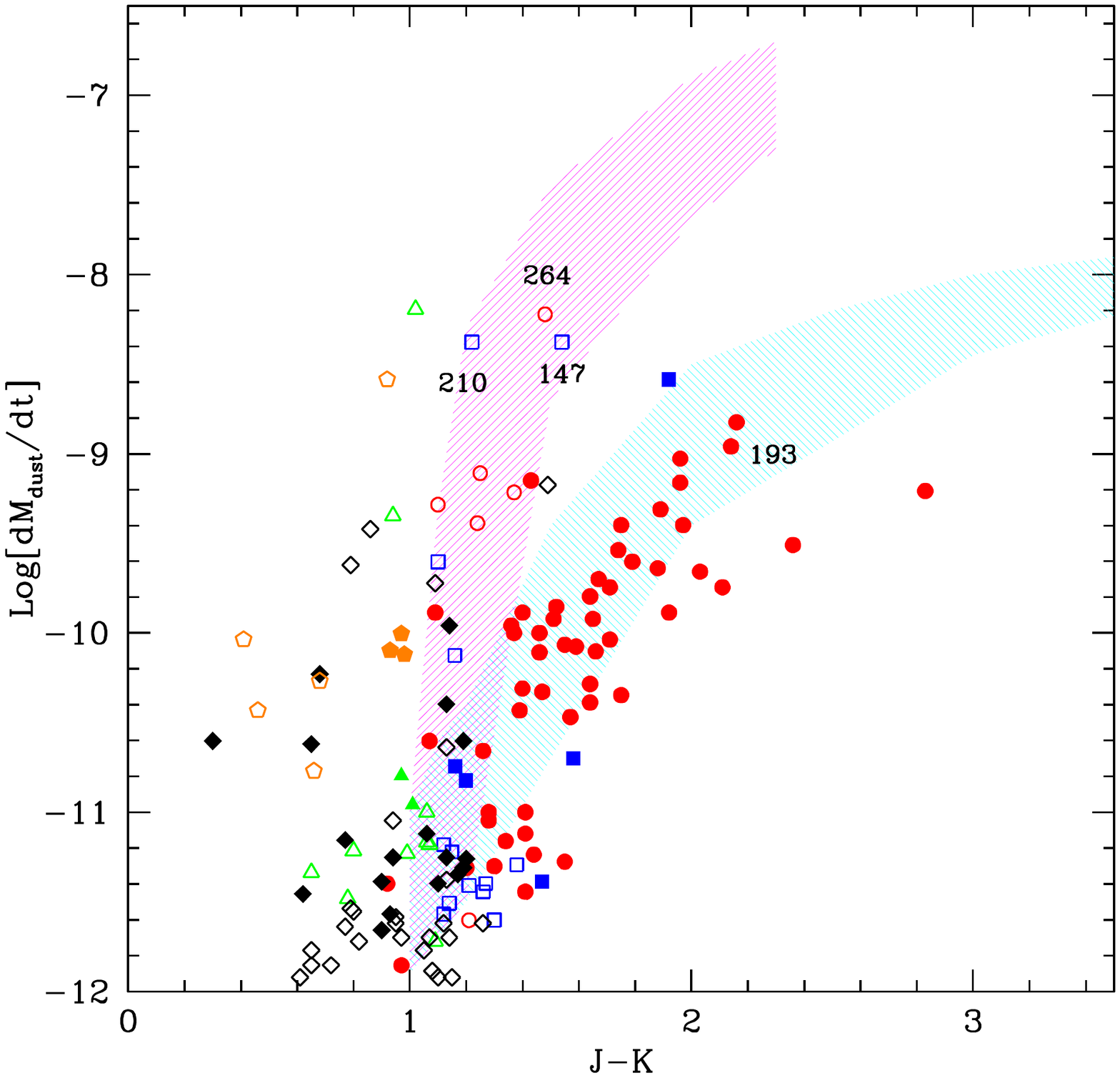}}
\end{minipage}
\vskip-50pt
\caption{The dust mass loss rates of Sextans A stars evaluated by \citet{jones18} as
a function of the $K-[4.5]$ (left) and $J-K$ (right) colours. The meaning of the 
different symbols is the same as in Fig.~\ref{foss}. The cyan and magenta regions indicate 
the expected dust mass loss rates of C-stars and HBB-stars, respectively, according to 
our modelling.}
\label{fdpr}
\end{figure*}

To this aim, we follow the method discussed in \citet{fg06} (see section 5.2), by which
the DPR of a star during a given evolutionary stage is calculated by means of the global
gas mass loss rate, the surface chemical composition and the fraction of gaseous molecules
condensed into dust particles. In this work we focus on the fraction of carbon and
silicon condensed into solid carbon and silicates, respectively. The contribution
of the rate of production of silicon carbide and of alumina dust are negligible in all
cases. For each evolutionary track the DPR increases with the amount
of dust present in the circumstellar envelope, which reflects into the definition of 
two different, tight relationships between the IR colours and the DPR, holding for C-stars 
and M-stars. Similarly to the IR fluxes in the different bands (see Fig.~\ref{foss}), we 
may identify the range of values expected for the DPR of C- and M-stars, as a function of 
the IR colours. This can be seen in Fig.~\ref{fdpr}, where we indicate with cyan and 
magenta shading the values of DPR expected for carbon and M stars, respectively. The 
vertical extension of the shaded regions at a given colour provides an estimate of the 
uncertainty associated to the theoretical DPR's. In the same figure we show the DPR of the 
individual sources, estimated by \citet{jones18}. In the comparison of the results obtained
in the present analysis with those by \citet{jones18}, we must consider that the methods
followed to estimate the individual DPR are completely different. While we followed
the approach described above, the estimates by \citet{jones18} are based on the procedure
of SED fitting. Therefore, despite the two analysis
are based on the same set of optical constants (i.e. Ossenkopf et al. 1992 for silicates,
Zubko et al. 1996 for solid carbon, Pegourie 1988 for SiC) it is not surprising that, as
we will see, for some stars the results obtained are significantly different.

For what concerns carbon stars, the largest DPR are from the stars with the reddest IR
colours, particularly the 14 sources on the red side of CMD1, discussed in the previous
section, with $K-[4.5] > 0.8$. These objects, for which we find 
$10^{-10}M_{\odot}/$yr $ < \dot M_d < 2\times 10^{-8}M_{\odot}/$yr, dominate the overall 
DPR of carbon stars.
A significant contribution to the DPR is provided by the stars ID 452, 462, 472,
which were classified as M-stars by \citet{jones18}, whereas according to our understanding 
they are obscured C-stars. For the other carbon stars, the range of values which we find
are in fine agreement with the estimates by \citet{jones18}. In particular, the stars ID 
193 and ID 454, classified as C-stars by \citet{jones18}, are expected to eject dust at
a rate $\dot M_d^{C} \sim 1-2 \times 10^{-9}M_{\odot}/$yr, in agreement with our
estimates. We believe that the two stars on the right region of the left panel of 
Fig.~\ref{fdpr} (ID 456 and 457) are C-stars, in agreement with \citet{jones18}; however,
based on their $(K-[4.5])$ colours, we find $\dot M_d \sim 3\times 10^{-8}M_{\odot}/$yr, 
a factor of 10 higher than \citet{jones18}.

Turning to M-stars, we find, in agreement with \citet{jones18}, that ID 147 and ID 210
are characterized by a large DPR, of the order of 
$\dot M_d \sim 4\times 10^{-9}M_{\odot}/$yr. We also consider 
ID 305, the brightest source within the M-stars obscuration sequence in the CMD2
(see Fig.~\ref{foss}), for which we estimate $\dot M_d \sim 3\times 10^{-8}M_{\odot}/$yr.
This star is extremely interesting, because the position on the CMD2 suggests
that it is a massive AGB star, descending from a $M>5~M_{\odot}$ progenitor. Unfortunately
no $[4.5]$ flux is available, thus we cannot confirm this conclusion on the basis of 
the CMD1 plane.
The sources ID 282 and 336 are also dusty M-stars, with 
$\dot M_d \sim 5\times 10^{-9}M_{\odot}/$yr. These 5 stars, other than the peculiar cases
which will be addressed in the following, provide the vast majority of the overall
DPR by M stars, which we find to be $\dot M_d^{M} = 4\times 10^{-8}M_{\odot}/$yr.

The source ID 17 was identified by \citet{jones18} as the one with the largest DPR and
was classified as an M star.
Unfortunately the $K$ flux is not available, thus we must base our interpretation on
the $[3.6]$ and $[4.5]$ fluxes. The tracks of M stars extend to $[3.6]-[4.5] \sim 0.6$,
significantly bluer than the colour of this star, which is $[3.6]-[4.5] \sim 1.2$. 
On the other hand, based on the position in the colour-magnitude  $([3.6]-[4.5], [4.5])$ 
plane, we interpret this object as a dusty C-star, descending from a $2-2.5~M_{\odot}$ 
progenitor, in the
final AGB phases, when most of the envelope was lost. Interestingly, despite this
interpretation is different from the M-star classification by \citet{jones18}, the
estimated DPR, $\dot M_d \sim 5\times 10^{-7}M_{\odot}/$yr, is very similar in the
two cases.

\begin{table*}
\caption{A summary of the main properties of the stars providing the dominant contribution
to the DPR of Sextans A. The different columns report the ID number according to the
catalogue by \citet{jones18} (1), whether the source is interpreted as C-star or M-star (2),
the estimated DPR (3), the range of mass of the progenitor (4), the current $C/O$ ratio (6)
and the $\log(C-O)+12$ (7), the classification according to \citet{jones18} (8) and the DPR 
estimated by \citet{jones18} (9).
}
\begin{tabular}{c c c c c c c c}        
\hline       
ID & SP & $\dot{M_d} (M_{\odot}$/yr) & $M_i (M_{\odot})$ & $C/O$ & $\log(C-O)+12$ & SPJ18 & 
$\dot{M_{d,J18}}(M_{\odot}$/yr) \\
\hline  
 452  &  C  & $(3\pm 1)\times 10^{-9}$    &   1.5-2.5  &  5-10    &  8.5-9.2 & M   &          -              \\
 462  &  C  & $(1\pm 0.5)\times 10^{-8}$  &   1.7-2.5  &  5-10    &  8.8-9.2 & M   &          -              \\
 472  &  C  & $(3\pm 1)4\times 10^{-9}$   &   2-2.5    &  5-10    &  8.5-9.2 & M   &          -              \\
 193  &  C  & $1-2\times 10^{-9}$         &   1.5-2.5  &  5-10    &  8.7-9.2 & C   & $1.1\times 10^{-9}$     \\
 454  &  C  & $1-2\times 10^{-9}$         &   1.7-2.5  &  5-10    &  8.5-9.2 & C   & $1.1\times 10^{-9}$     \\
 456  &  C  & $(3\pm 1)\times 10^{-8}$    &   2-2.5    &  5-10    &  9-9.2   & C   & $4\times 10^{-9}$       \\
 457  &  C  & $(3\pm 1)\times 10^{-8}$    &   2-2.5    &  5-10    &  9-9.2   & C   & $2.3\times 10^{-9}$     \\
 147  &  M  & $(3\pm 1)\times 10^{-9}$    &   3.5-5    &  0.1-0.3 &    -     & M   & $4.2\times 10^{-9}$     \\
 210  &  M  & $(3\pm 1)\times 10^{-9}$    &   3.5-5    &  0.1-0.3 &    -     & M   & $4.2\times 10^{-9}$     \\
 305  &  M  & $(3\pm 1)\times 10^{-8}$    &   5-7.5    &  0.1-0.3 &    -     & -   &          -              \\
 282  &  M  & $2-10\times 10^{-9}$        &   4-5      &  0.1-0.3 &    -     & -   &          -              \\ 
 336  &  M  & $2-10\times 10^{-9}$        &   4-5      &  0.1-0.3 &    -     & -   &          -              \\
 17   &  C  & $5\times 10^{-7}$           &   2-3      &  5-10    &    -     & M   & $5.5\times 10^{-7}$     \\
\hline       
\label{tabdpr}
\end{tabular}
\end{table*}

The characterization of the source ID 264 is also cumbersome, given the lack of the 
$[4.5]$ data, which leaves the position on the CMD2 as the only possibility to characterize
this object. It is classified as M by \citet{jones18} at odds with our interpretation,
based on the position on the CMD2, which falls within the region populated by C-stars.
We estimate a small DPR of the order of $\dot M_d \sim 5\times 10^{-10}M_{\odot}/$yr,
significantly different from the estimate by \citet{jones18}, which include ID 264
among the stars with the highest DPR.

Finally, we consider ID 448. In this case we lack the J magnitude, thus we may only rely
on the analysis of the CMD1. The star is located within the C-star sequence, at 
$K-[4.5] \sim 1$. According to our analysis the DPR of this star is 
$\dot M_d \sim 3\times 10^{-10}M_{\odot}/$yr, significantly smaller (by a factor $\sim 20$) 
than predicted by \citet{jones18}.

The results of the previous discussion, limited to the stars providing most of the
contribution to the global DPR of Sextans A, are reported in table \ref{tabdpr}. For each 
of the sources considered we give the estimated DPR according to the present analysis,
the range of mass of the progenitor and the current $C/O$ surface ratio.

Based on these results, we find that
the overall DPR for Sextans A is $\dot M_d = 6\times 10^{-7}M_{\odot}/$yr, distributed
among C-stars, with $\dot M_d^{C} = 5.6\times 10^{-7}M_{\odot}/$yr, and M stars, with
$\dot M_d^{M} = 4\times 10^{-8}M_{\odot}/$yr. 
The overall DPR given above is in excellent
agreement with the results from \citet{jones18}, whereas it is a factor $\sim 6$ higher than
\citet{boyer15b}. 

A possible source of uncertainties to the above
estimate is associated to the stars for which we could not provide any reasonable
classification (indicated with open and full diamonds in Fig.~\ref{foss} and \ref{fdpr}),
and for the stars whose position in CMD1 and/or CMD2 is at the edge between the M-stars
and the C-stars region, whose classification has some degree of arbitrariness (see discussion
in section \ref{distance}). For the first group of stars we estimate that the overall DPR 
is $1.3\times 10^{-8}M_{\odot}/$yr. For the second sample of objects, the estimated DPR 
is $\sim 10^{-8}M_{\odot}/$yr. Even in the unlikely case that all these sources have been 
misclassified, the overall DPR would remain practically unchanged, owing to the significant 
overlapping of the cyan and magenta zones in Fig.~\ref{fdpr}, which means that no 
substantial difference is expected in the DPR of C- and M-stars.

\section{The role of a metallicity spread}
\label{zeta}
The analysis presented in the previous sections was based on the assumption that
the stars in Sextans A share the same metal-poor chemical composition, corresponding to
a metallicity $Z=10^{-3}$. Because we consider probable the presence of a metallicity
spread, we believe important to discuss how the conclusions reached so far might be
affected by a change in metallicity. To this aim, we consider models of metallicity
$Z=3\times 10^{-4}$ and $Z=2\times 10^{-3}$. The latter models were presented in
\citet{ventura16a}. For the $Z=3\times 10^{-4}$ stars of initial mass above
$2.5~M_{\odot}$ we used the models by \citet{marcella13}; in the low mass domain
we calculated ad hoc models specifically for the present investigation.

\begin{figure*}
\begin{minipage}{0.48\textwidth}
\resizebox{1.\hsize}{!}{\includegraphics{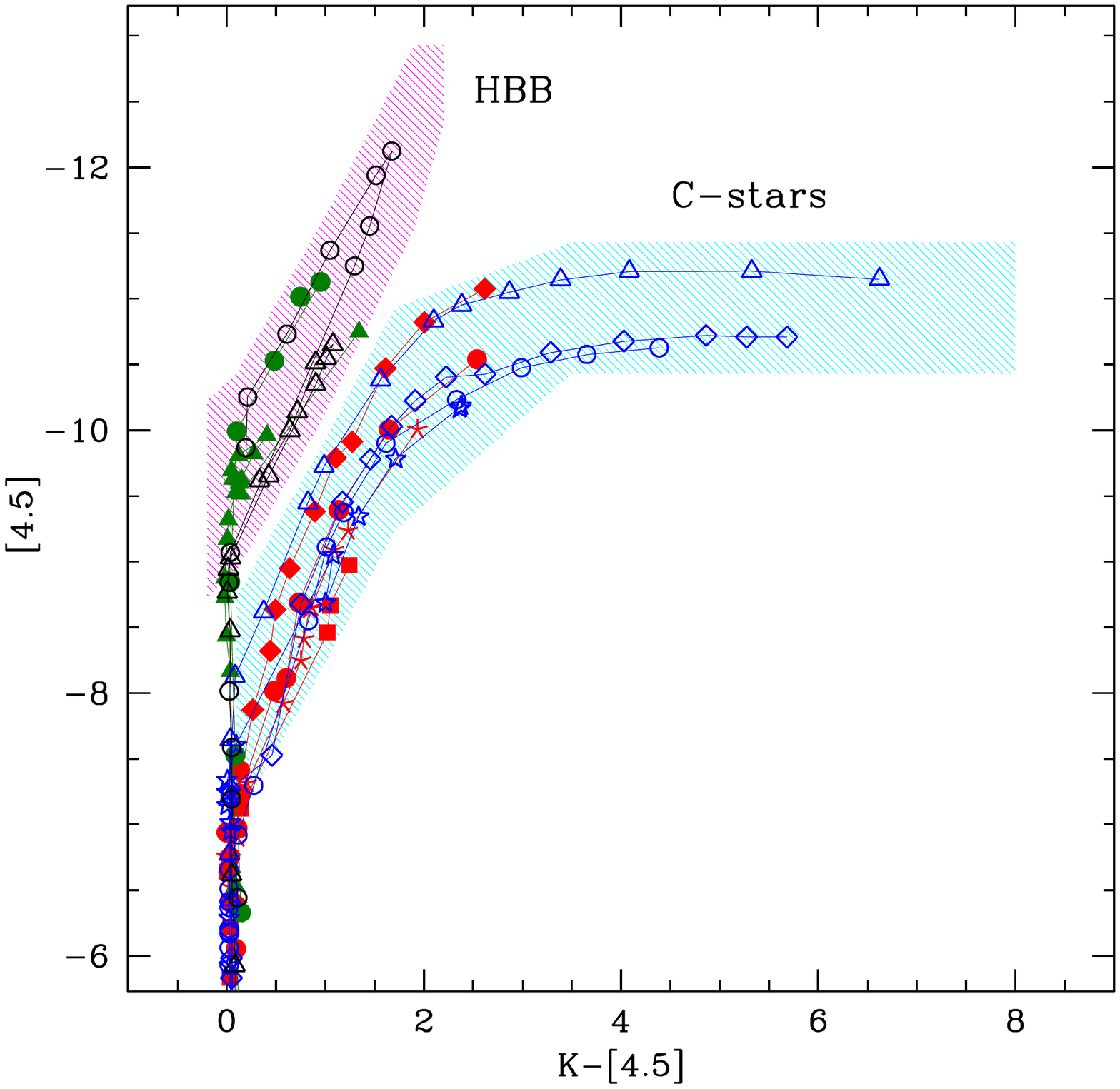}}
\end{minipage}
\begin{minipage}{0.48\textwidth}
\resizebox{1.\hsize}{!}{\includegraphics{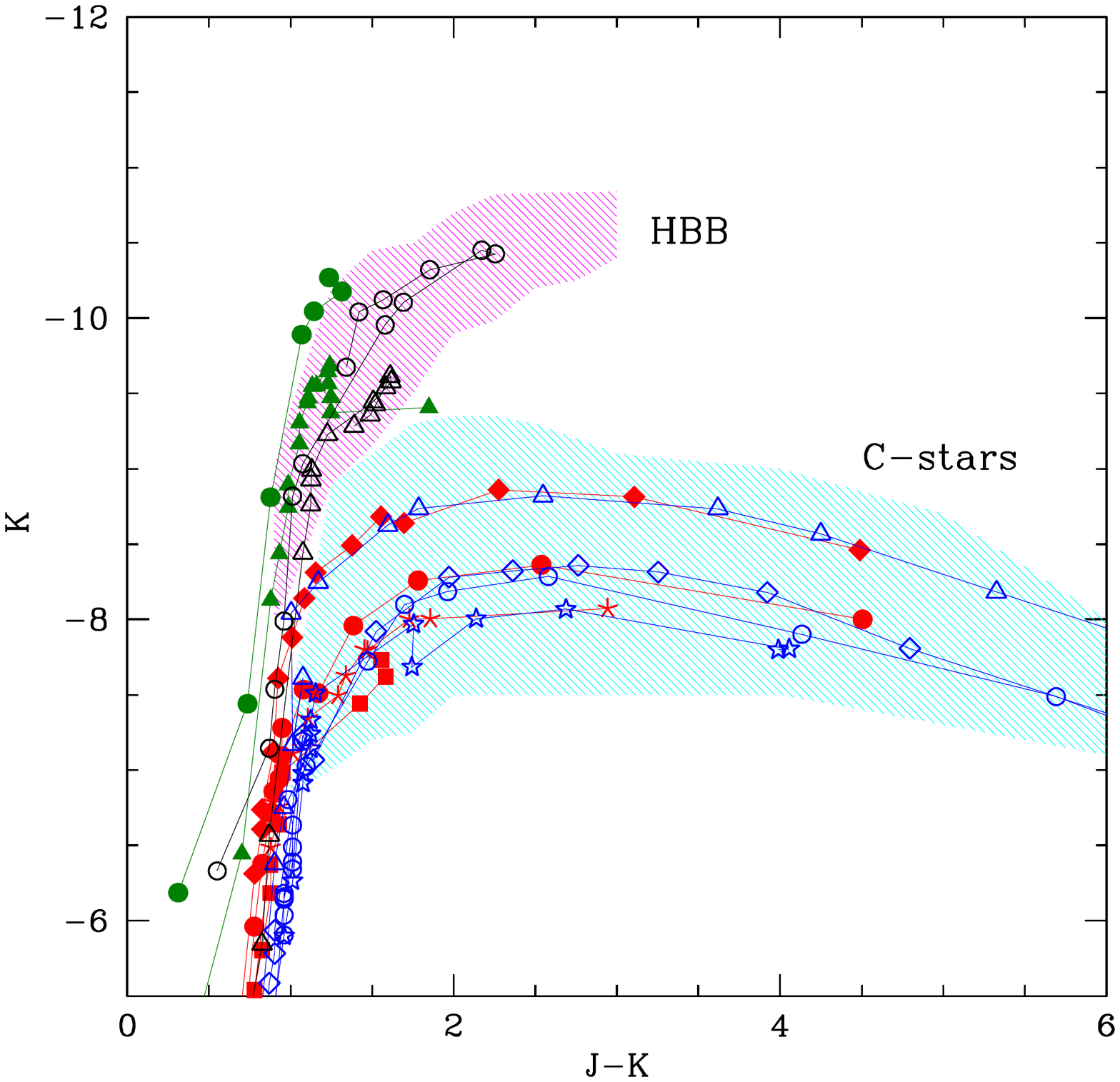}}
\end{minipage}
\vskip-50pt
\caption{The evolutionary tracks of stars of metallicity $Z=3\times 10^{-4}$ and
$Z=2\times 10^{-3}$ in the CMD1 (left panel) and CMD2 (right) planes. In the low-mass
domain we indicate with red, full points the $Z=3\times 10^{-4}$ models and
with blue, open points their $Z=2\times 10^{-3}$ counterparts, with the following
symbols: $1.1~M_{\odot}$ - squares; $1.3~M_{\odot}$ - asterisks; $1.5~M_{\odot}$ - 
circles; $2~M_{\odot}$ - diamonds; $2.5~M_{\odot}$ - triangles. For the stars experiencing
HBB we indicate with black, open triangles (circles) the tracks of $4~M_{\odot}$ 
($6~M_{\odot}$) stars of metallicity $Z=2\times 10^{-3}$, whereas green, full
triangles and circles indicate their $Z=3\times 10^{-4}$ counterparts.}
\label{fzeta}
\end{figure*}

Fig.~\ref{fzeta} shows the tracks of stars of different mass in the CMD1 and CMD2
planes. The regions where C-stars and obscured M-stars are expected to evolve have
been highlighted, respectively, with cyan and magenta shading, similarly to
Fig.~\ref{ftracks} and \ref{foss}.

Regarding obscured C-stars, we find that the magnitudes and the magnitude spread are
approximately independent of metallicity; this holds for both the CMD1 and CMD2 planes.
This result is motivated by the dominant role played by the core mass in determining the
main evolution properties of AGB stars, particularly the extent of the TDU and the
ignition and the strength of HBB; because the core mass directly affects the luminosity 
of the star, it is not surprising that stars in the same evolutionary stage, sharing the
same chemical and dust properties, evolve at similar magnitudes.

While the expected IR fluxes of obscured C-stars are fairly independent of the chemical
composition, Fig.~\ref{fzeta} shows that the metallicity has an effect on the extension 
of the obscuration sequence, which is shorter the lower is $Z$. This result, 
discussed in \citet{ventura14}, is due to the higher effective temperatures and smaller 
radii of lower-Z stars, which favour lower rates of mass loss and of dust grains growth.

The sources in the sample used here extend to $K-[4.5] \sim 3$ and $J-K \sim 3$ in the
CMD1 and CMD2 planes, respectively. This is of little help for an evaluation of the
metallicity, because the theoretical sequences reach such IR colours for all the
$Z$'s considered. The only exception are ID 456, 457 and 462, for which we 
rule out that $Z < 10^{-3}$. 

The absence of stars with $K-[4.5] > 3$ cannot be used as a metallicity discriminator 
either, because the sample adopted here is limited and those evolutionary phases are
extremely rapid, with little possibility that a star is detected (see e.g. Fig.~2 in
Dell'Agli et al. (2018)).

While the present analysis shows that this methodology does not allow a robust evaluation 
of the metallicity of the sources observed, a significant result is that the determination 
of the distance of the galaxy based on the magnitudes of the C-stars sequence, as described 
in section \ref{distance}, is independent of $Z$. The same holds for the determination of 
the DPR, which is essentially related to the IR colours.

For what attains M-stars, we find that little difference exists between the $Z=10^{-3}$
and $Z=2\times 10^{-3}$ models, in terms of the extension of the tracks in the CMD1 and
CMD2 planes. This is because the higher the metallicity the weaker the strength of HBB
(see e.g. Ventura et al. 2013), thus the rate of dust grains formation; this effects
counterbalances the higher availability of silicon in the surface regions of the star.
The extension of the $Z=3\times 10^{-4}$ tracks appear not compatible with the 
$J-K$ colours of the most obscured M stars, for which we may claim a metallicity 
$Z \geq 10^{-3}$.

\section{Conclusions}
\label{conclusions}
We study the evolved stellar population of the metal-poor galaxy Sextans A, adopting the 
theoretical description of the structural and evolutionary properties of intermediate mass 
stars evolving through the AGB phase and of massive stars during the core He-burning stage.
The evolutionary tracks in the observational planes obtained with IR fluxes are compared
with results from NIR and mid-IR photometry. In particular, we base our analysis on the
sources included in a sample of stars in Sextans A, presented recently by \citet{jones18}.

The simultaneous analysis of the position of the stars in the $(K-[4.5], [4.5])$ and
$(J-K, K)$ planes allowed the identification of $\sim 90$ C-stars, descending from
$1-3~M_{\odot}$ stars, formed between 500 Myr and 6 Gyr ago. 14 out of these objects
are heavily obscured and are interpreted as stars evolving through the late evolutionary 
phases, surrounded by large quantities of dust, mainly composed of solid carbon 
particles; we find that the size of these dust grains in the range $0.1-0.2~\mu$m, but
this result is more uncertain, as it is sensitive to the assumptions made regarding the
number density of seed particles available. These stars provide the vast majority of
the overall DPR from C-stars, which is of the order of $\sim 5.6\times 10^{-7} M_{\odot}/$ yr.

We also identified $\sim 30$ stars currently undergoing HBB, descending from $M \geq 4~M_{\odot}$
progenitors. The presence of these objects confirms that significant star formation took 
place in Sextans A during the last 200 Myr. According to our interpretation these 
stars are surrounded by silicate particles of $0.05-0.07 ~\mu$m size and eject into their 
surroundings dust with a rate of $\sim 4\times 10^{-8} M_{\odot}/$ yr. These results
need confirmation from hydrodynamical simulations, because silicate particles of similar
size, when applied to low-mass stars, can hardly drive a wind.

This study confirms that the combined use of NIR and mid-IR data allow to classify evolved
stars, although a more robust classification will be possibile with the future JWST
observations. The results regarding dust production confirm that significant amount of
dust is produced by AGB stars in metal-poor environments, with a dominant contribution
provided by carbon stars.

\section*{Acknowledgments}
FDA and DAGH acknowledge support provided by the Spanish Ministry of
Economy and Competitiveness (MINECO) under grant AYA-2017-88254-P. The authors
are indebted to the anonymous referee for the careful reading of the manuscript
and for the several comments and suggestions, which helped to improve the quality
of the manuscript.

\end{document}